\documentclass[12pt,preprint]{aastex}

\usepackage{epsfig}

\begin{document}

\title{Turbulence and Dynamo in Galaxy Cluster Medium: \\
  Implications on the Origin of Cluster Magnetic Fields }


\author{Hao Xu\altaffilmark{1,2}, 
Hui Li\altaffilmark{2},  
David C. Collins\altaffilmark{1},  
Shengtai Li\altaffilmark{2},
and Michael L. Norman\altaffilmark{1}}
\altaffiltext{1}{Center for Astrophysics and Space Sciences,
  University of California, 
San Diego, 9500 Gilman Drive, La Jolla, CA 92093; haxu@ucsd.edu, dcollins@physics.ucsd.edu, mlnorman@ucsd.edu}
\altaffiltext{2}{Theoretical Division, Los Alamos National Laboratory, Los
  Alamos, NM 87545; hli@lanl.gov, sli@lanl.gov}

\begin{abstract}
  We present self-consistent cosmological magnetohydrodynamic (MHD)
  simulations that simultaneously follow the formation of a galaxy
  cluster and the magnetic field ejection by an active galactic nucleus
  (AGN).  We find that the magnetic fields ejected by the AGNs, though
  initially distributed in relatively small volumes, can be
  transported throughout the cluster and be further amplified by the
  intra-cluster medium (ICM) turbulence during the cluster
  formation process.  The ICM turbulence is shown to be generated and sustained by
  the frequent mergers of smaller halos. Furthermore, a cluster-wide dynamo process
  is shown to exist in the ICM and amplify the magnetic field energy
  and flux. The total magnetic energy in the cluster can reach $\sim$
  $10^{61}$ ergs while micro Gauss ($\mu$G) fields can distribute over
  $\sim$ Mpc scales throughout the whole cluster. This finding shows
  that magnetic fields from AGNs, being further amplified by the ICM
  turbulence through small-scale dynamo processes, can be the origin
  of cluster-wide magnetic fields.
\end{abstract}

\section{Introduction}

There is growing evidence that the ICM is permeated with magnetic
fields, as indicated by the detection of large-scale, diffused radio
emission called radio halos and relics \citep[see recent reviews by][]{Ferrari08,Carilli02}.
The radio emissions are extended over $\ge 1$ Mpc, covering the whole cluster. By assuming that
the total energy in relativistic electrons is comparable to the
magnetic energy, one often deduces that 
the magnetic fields in the cluster halos  
can reach $0.1-1.0$ $\mu$G and the total magnetic energy can be
as high as $10^{61}$ ergs \citep{Feretti99}.  The Faraday rotation
measurement (FRM), combined with the ICM density measurements, often
yields cluster magnetic fields of a few to ten $\mu$G level (mostly in
the cluster core region). More interestingly, it reveals that magnetic
fields can have a Kolmogorov-like turbulent spectrum in the cores of
clusters \citep{Vogt03} with a peak at several kpc. Other studies have 
suggested that the coherence scales of magnetic fields can range from a 
few kpc to a few hundred kpc \citep{Eilek02, Taylor93, Colgate00}, implying
large amounts of magnetic energy and fluxes. Understanding the origin
and effects of magnetic fields in clusters is important because they
play a crucial role in determining the structure of clusters through
processes such as heat transport, which consequently affect the
applicability of clusters as sensitive probes for cosmological
parameters \citep{Voit05}.

Although the existence of cluster-wide magnetic fields is clear, their
origin is still poorly understood.  Two scenarios have received most
attention: a) magnetic fields are initially from the outflows of
normal or active galaxies \citep{Donnert08, Furlanetto01, Kronberg01},
and such fields can be further amplified by cluster merger
\citep{Roettiger99} and turbulence \citep{Dolag02, Dubois08}; and b)
very small proto-galactic seed fields are amplified by dynamo
processes in clusters \citep{Kulsrud97,Ryu08}, though the exact
mechanism for dynamo is still being debated \citep{Bernet08}.  Large
scale radio jets from AGNs serve as one of the most intriguing
candidates in the first scenario because they could carry large amount
of magnetic energy and flux
\citep{Burbidge59,Kronberg01,Croston05,McNamara07}.  The magnetization
of the ICM and the wider inter-galactic medium (IGM) by AGNs has been
suggested on the energetic grounds
\citep{Colgate00,Furlanetto01,Kronberg01}, though the exact physical
processes of how magnetic fields will be transported and amplified
remain sketchy. Some of the key physics questions in these models
include: what is the volume filling process of AGN magnetic fields in
the ICM/IGM? Is the ICM turbulent and what are its properties? Is
there a dynamo in the ICM that can amplify fields?

In this Letter, we present self-consistent cosmological MHD
simulations to address the question of the origin of magnetic fields
in clusters. We explore specifically the scenario that cluster-wide
magnetic fields initially came from the magnetic fields of an AGN. We
also describe the properties of the ICM turbulence and demonstrate the
existence of dynamo in the ICM. The simulations are described in
Section \ref{sec:model} and results are presented in
Section \ref{sec:result}. Discussions are given in Section 4.

\section{Basic Model and Simulations}
\label{sec:model}

We have performed detailed cosmological MHD simulations of galaxy
cluster formation with the magnetic field injection from an AGN, using
the newly developed ENZO+MHD code, which is an Eulerian cosmological
MHD code with adaptive mesh refinement (AMR) \citep{Xu08a}. Our
simulations use a $\Lambda$CDM model with parameters $h=0.7$,
$\Omega_{m}=0.3$, $\Omega_{b}=0.026$, $\Omega_{\Lambda}=0.7$, and
$\sigma_{8}=0.928$, with initial conditions extracted from the
Simulated Cluster Archive (see http://lca.ucsd.edu/data/sca/).
The simulation volume is $366$ Mpc on a side, and it uses a $128^3$
root grid and $2$ level nested static grids in the Lagrangian region
where the cluster forms. This gives an effective root grid resolution
of $512^3$ cells ($0.5$ Mpc) and dark matter particles of mass $1.49
\times 10^{10}M_{\odot}$.

The simulations were evolved from redshift $z=30$ to $z=0$
adiabatically.  We ``turn on'' the AGN magnetic injection at redshift
$z=3$, centered at a massive galaxy in a proto-cluster which has a
virial radius $r_v \approx 0.15$ Mpc, a baryon virial mass $m_b
\approx 3.2 \times 10^{11}~M_{\odot}$, and a virial total mass $m_v
\approx 7.6 \times 10^{12}~M_{\odot}$. The cluster eventually grows to
$r_v \approx 2.15$ Mpc, $m_b \approx 9.0 \times 10^{13}~M_{\odot}$,
and $m_v \approx 1.1 \times 10^{15}~M_{\odot}$ by $z=0$. The total injected
magnetic energy by AGN is about $2 \times 10^{60}$ ergs, with an
average input power of $1.75 \times 10^{45}$ ergs s$^{-1}$ for a
duration of $36$ Myr. The magnetic energy is injected inside $0.2$
$r_v$.  Because it is currently not possible to resolve both the galaxy cluster and the AGN environment simultaneously, we have adopted an approach that mimics the possible magnetic energy injection by an AGN \citep{Li06}.
The size of the injection region and the associated field strength are not realistic when compared to the real AGN jets, but on global scales, the previous studies by \citet{Nakamura06} and \citet{Xu08a} showed that this approach can reproduce the observed X-ray bubbles and shock fronts \citep{McNamara05}. 
AMR is allowed only in a region of (50 Mpc)$^3$ where the galaxy cluster forms. During the cluster formation, the refinement is controlled by baryon and dark matter overdensity. In addition, all the region where magnetic field strength is higher than $10^{-7}$ G will be refined to the highest level. 
There are a total of $8$ levels of refinement beyond the root grid,
for a maximum spatial resolution of $11.2$ kpc.  Consequently, our
simulations are equivalent to $\sim 600^3$ uniform grid MHD runs in
the cluster region with full cosmology. The simulation was performed
on the linux cluster Coyote at LANL with about 300,000 CPU hours consumed.

\section{Results}
\label{sec:result}

\subsection{Global Morphology}

To illustrate the formation and evolution of the cluster along with
the evolution of magnetic fields from the AGN, we present images of
the projected gas density and magnetic energy density at different
stages of cluster formation in Fig. \ref{fig:density_med}.  At $z=3$
($t= 0$), an AGN is ``turned on'' in a sub-cluster (which eventually
merges with another sub-cluster about $400$ Myr after the AGN
injection has finished). The AGN is ``turned off'' at $t=36$ Myr
($z=2.95$). At $t=180$ Myr, we see the density cavities produced by
the AGN magnetic fields, reminiscent of the jet-lobe structure of
radio galaxies \citep{Xu08a}.  At a later time $t=468$ Myr, the
jet-lobe structure is destroyed by the cluster mergers. When two
sub-clusters finish merging at about $t=1.1$ Gyr, a large part of
magnetic fields is carried out of the cluster center region by the
ejected flow from mergers. As the evolution progresses, magnetic
fields, which follow the plasma motion, are being sheared, twisted,
and spread throughout the whole cluster.  Judging by the images from
$t=1.1$ to $t=6.28$ Gyr, this volume-filling process is quite
efficient (see Discussions). 
By $z=0.5$ ($t=6.28$ Gyr), magnetic fields are well mixed
with the ICM and are distributed throughout the whole
cluster, with some high magnetic field regions obviously from shock
compressions.
At $z=0.0$, while the cluster has relaxed, the magnetic fields seem
to distribute over the whole cluster quasi-uniformly. 

\subsection{Energy Evolution and Magnetic Field Radial Profile}

\begin{figure}
\begin{center}
\epsfig{file=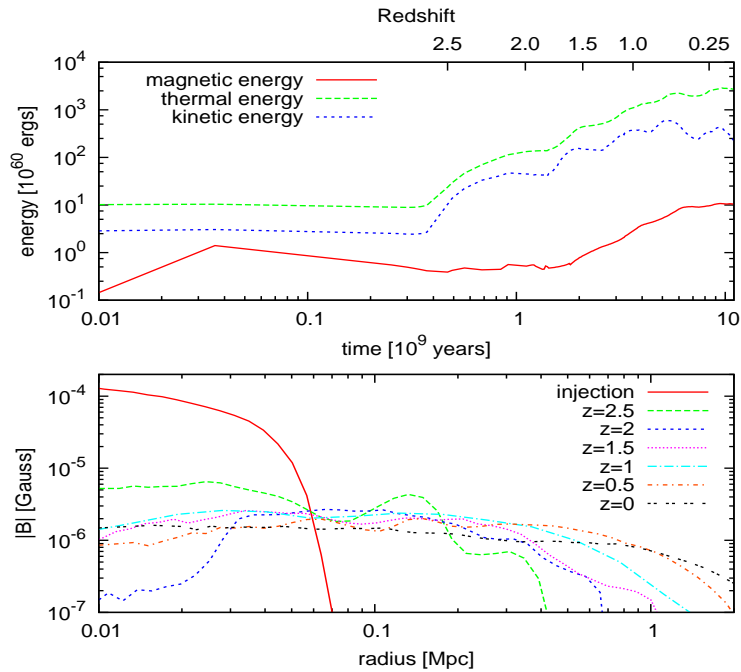,height=3.5in,width=4in} 
\end{center}
\caption{Top panel: Temporal evolution of different components of energy inside
  the virial radius of the cluster. The variations in the thermal and
  kinetic energies are due to mergers. Bottom panel: The spherically averaged
  radial profile of magnetic field strength at different epochs of the
  cluster formation. The radius is measured in the proper frame. They
  show the turbulent diffusion of magnetic fields throughout the
  cluster, yet maintaining its strength via the dynamo process.
\label{fig:energy}}
\end{figure}

The evolution of the total thermal, kinetic, and magnetic energy
inside the cluster's virial volume is shown in the top panel of Fig.
\ref{fig:energy}. The kinetic energy is calculated as the turbulent
component by subtracting the bulk flow motion. By the AGN injection,
$\sim 2\times 10^{60}$ ergs of magnetic energy is input into a
sub-cluster. A significant fraction ($\sim 80\%$) of this energy is
deposited into the ICM due to expansion and heating \citep{Xu08a}. When
the two big sub-clusters merge at $t=400$ Myr, the total thermal and
kinetic energies of the cluster increase by ten-fold by $t \sim 600$
Myr. During this major merger, the magnetic fields are still very
local, largely in magnetic bubbles. So the merger event itself did not
significantly change the magnetic field energy.  The thermal and
kinetic energies of the cluster continue to increase as the cluster
grows in mass via accretion and mergers as indicated by the variations
in the energy evolution curves. At $t\sim 2$ Gyr, the magnetic fields
from the AGN have been spread throughout a significant volume of the
cluster (see Fig. \ref{fig:density_med}). Starting from this time, the
magnetic energy experiences an exponential increase by a factor of
$20$ until $t\sim 6$ Gyr, at which time the growth has slowed. After
$t \sim 6$ Gyr, the cluster grows slowly in its total energy and
becomes relaxed, and the magnetic energy increases slowly with the
growth of the cluster. By $z = 0$, the total magnetic energy has
reached $\sim 10^{61}$ ergs inside the cluster. Considering that the
magnetic energy will drop as $\propto r^{-1}$ due to the the expansion
of the universe, the total magnetic energy has actually increased by
$\sim 75$ times between $t=2$ and $t=10$ Gyr.  

In the bottom panel of
Fig. \ref{fig:energy}, we present the spherically averaged radial
profiles of magnetic field strength at different epochs. 
At $z=0$, the magnetic field
strength is $\sim 1.5$ $\mu$G at the core,
decreasing slowly to  $\sim 0.7$ $\mu$G at $\sim$ Mpc radius.
These radial profiles are different from what is shown by other simulations
which often exhibit faster radial decline \citep{Dolag02,Dubois08, Donnert08}. 
It is presently unclear whether this difference is caused by the
different origins of ``seed'' magnetic fields or by the effects of
different numerical techniques and resolutions. This difference
deserves further study since it may be used to distinguish different
origins of magnetic fields in clusters. Furthermore, it will be
necessary to compare our simulation results with cluster magnetic
observations in detail \citep[e.g.,][]{Govoni06,Guidetti08}. This will
be presented in future publications. 

\subsection{Small-scale Turbulent Dynamo and MHD Turbulence}

\begin{figure}
\begin{center}
\epsfig{file=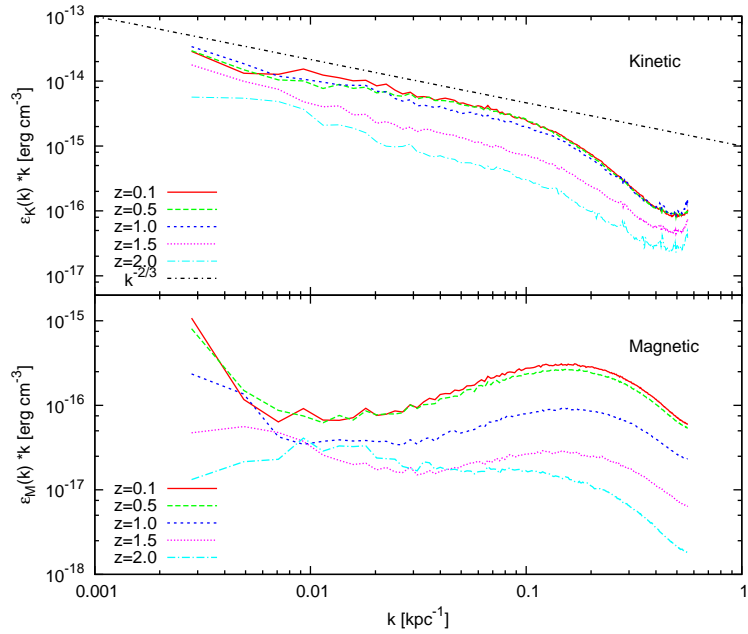,height=3.5in,width=4.0in} 
\end{center}
 \caption{Power spectra of the kinetic energy density and magnetic
   energy density of the ICM at different epochs. The ICM turbulence
   is represented by the Kolmogorov-like spectra in kinetic energy.
   The magnetic energy is amplified via a dynamo process.
   \label{fig:power}}
\end{figure}

The physical origin for the exponential amplification of the magnetic
energy is due to a cluster-wide turbulent dynamo process. In
Fig. \ref{fig:power}  we present the power spectra of the ICM plasma's
kinetic  energy density and magnetic energy density in a comoving cube
with  $5.71$ Mpc on the side. The scales are shown in the comoving
units so that  they are not affected by the universe expansion and the
power  changes are from the dynamics of cluster evolution alone. Here,
$k = 0.003$ kpc$^{-1}$ corresponds to the maximum radius of the
central  region $2.86$ Mpc where a high spatial refinement is adopted.
Since the cluster's total thermal energy is a factor of $\sim 3-5$
larger than  its kinetic energy, the ICM can be thought as nearly
incompressible  plasmas overall, though shocks generated by mergers
are very  frequent and important. In fact, the flows and shocks from
mergers  tend to be global and propagate through the whole cluster
\citep[see also  the earlier work by][]{Roettiger99}.  We can divide
the full  kinetic spectrum approximately into three regions: the
``injection''  region for $k \sim 0.003 - 0.01$ kpc$^{-1}$ where the
large scale  flows and shocks from the mergers ``stir'' the whole
cluster; the  ``cascade'' region for $k \sim 0.01 - 0.1$ where the
spectrum shows  a smooth power law similar to a Kolmogorov-like
incompressible  turbulence; the ``dissipation'' region for $k > 0.1$
where the  spectrum steepens gradually. We attribute this steepening
to both  the dissipation by shocks (which has a physical origin) and
the limited  spatial resolution (which has a numerical origin). These
features  in the kinetic energy density spectrum lead us to conclude
that the  ICM is turbulent and our simulations have captured the
essence of this  turbulence. This ICM turbulence is in a
driven-dissipative state  where frequent mergers will drive the
turbulence over  relatively short time scales (a few Myr) but the
turbulence is  decaying in-between mergers on timescales of $\sim$ Gyr
(which  is approximately the dynamic timescale for the whole cluster).

\begin{figure}
\begin{center}
\epsfig{file=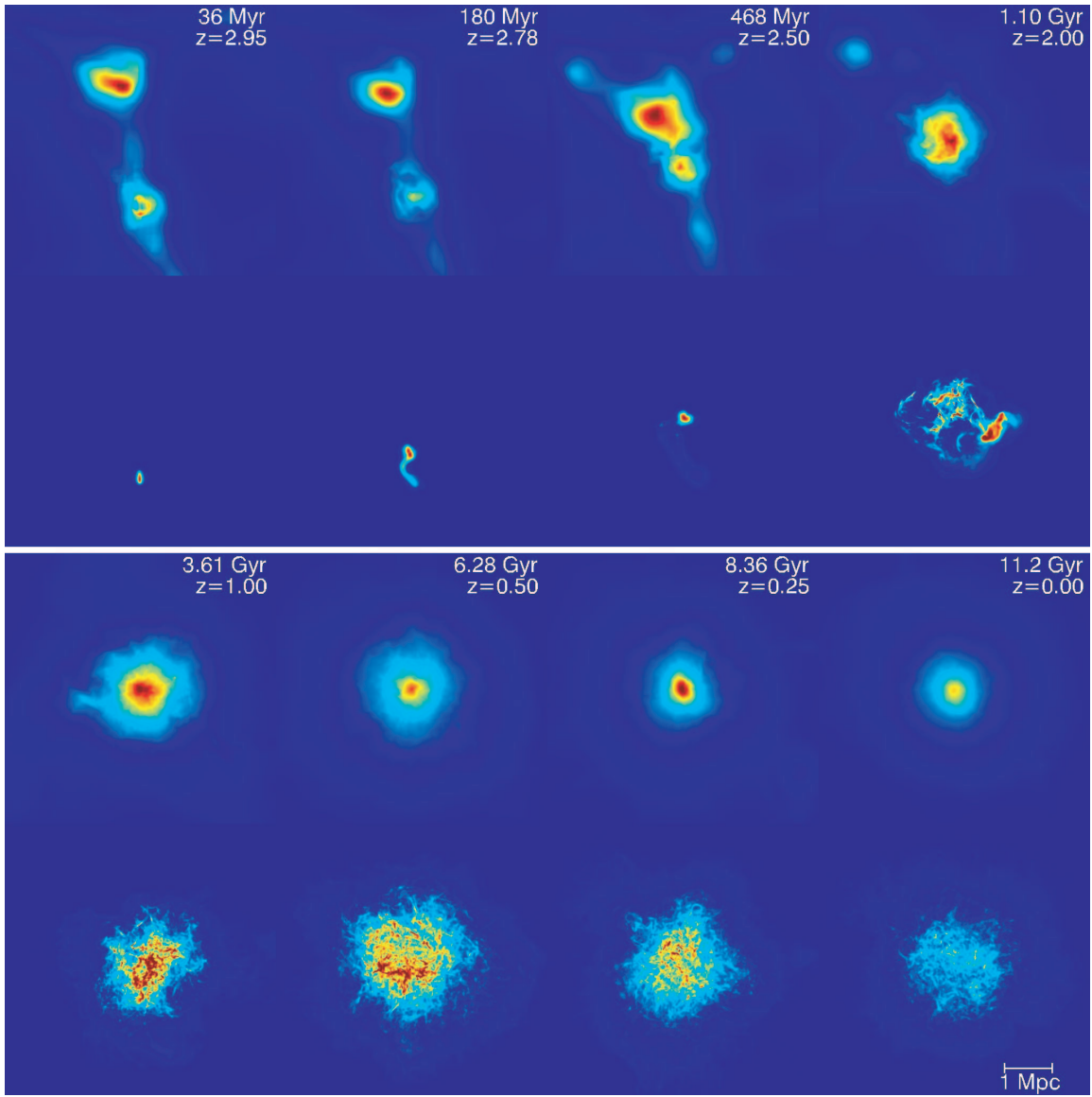,height=6.035in,width=6in}
\end{center}
\caption{Snap shots of the projected baryon density (upper rows) and
  magnetic energy density (lower rows) for different epochs of cluster 
  formation and evolution. Each image covers a region of
  $5.71$ Mpc $\times$ $5.71$ Mpc (comoving). The projected results
  are obtained by integrating $5.71$ Mpc (comoving) centered at
  the cluster along lines perpendicular to the observed plane. The eight columns are marked
  with the time $t$ since the AGN injection and the respective
  redshift $z$. The top panel uses different color scale for each plot
  to have the best visual effect. The color range of the bottom panel
  is the same for all subplots, ranging
  from $4.2 \times 10^{19}$ to  $1.3 \times 10^{22}$
  cm$^{-2}$ for the baryon particle number density and from $0$ to
  5 $\times$ $10^{11}$ ergs cm$^{-2}$ for the integrated magnetic energy density.
\label{fig:density_med}}
\end{figure}

The magnetic energy density spectrum also has three regions in the
$k-$space corresponding to the kinetic energy spectrum. From $t=2$ to
$t=10$ Gyr ($z=1.5$ to $z=0.1$), the magnetic spectrum for $k \sim
0.01 - 0.6$ retains an ``invariable'' shape but the energy density
undergoes exponential amplification then goes into saturation. This is
a strong signature for the so-called small-scale turbulent dynamo \citep{Brandenburg05}.
Furthermore, the magnetic energy density peaks at $k \sim 0.2$ with
$\sim 3\times 10^{-16}$ ergs cm$^{-3}$. The corresponding kinetic
energy density is $\sim 8 \times ~10^{-16}$ ergs cm$^{-3}$. So, the
magnetic energy for $k > 0.1$ has saturated at a sub-equipartition
level (by a factor of $\sim 3$). The drop-off at high $k$ should be
due to the numerical dissipation. For $k < 0.01$, neither kinetic or
magnetic energy seems to have saturated. It is interesting to note
that the dynamo process starts vigorously only at $t \sim 2$ Gyr
($z=1.5$), when the magnetic fields have been spread through a
significant fraction of the whole cluster (see Fig.
\ref{fig:density_med}). Putting Figs.  \ref{fig:density_med} -
\ref{fig:power} together, we see that the ICM turbulence both
amplifies the magnetic energy and diffuses the fields throughout the
cluster. The magnetic energy density saturation occurred at $t\sim 6$
Gyr but magnetic fields continue to ``grow'' in their spatial extent
through turbulent diffusion.  Furthermore, the un-signed magnetic flux
through the mid-plane of the cluster has exponentially grown from $7.6
\times 10^{41}$ to $1.23 \times 10^{43}$ G cm$^2$ from $z=2$ to
$z=0.5$, which is another clear indication of the turbulent dynamo
that is responsible for both amplifying the field energy and diffusing
the field through the cluster.

\subsection{Faraday Rotation Measurement}
\begin{figure}
\begin{center}
\epsfig{file=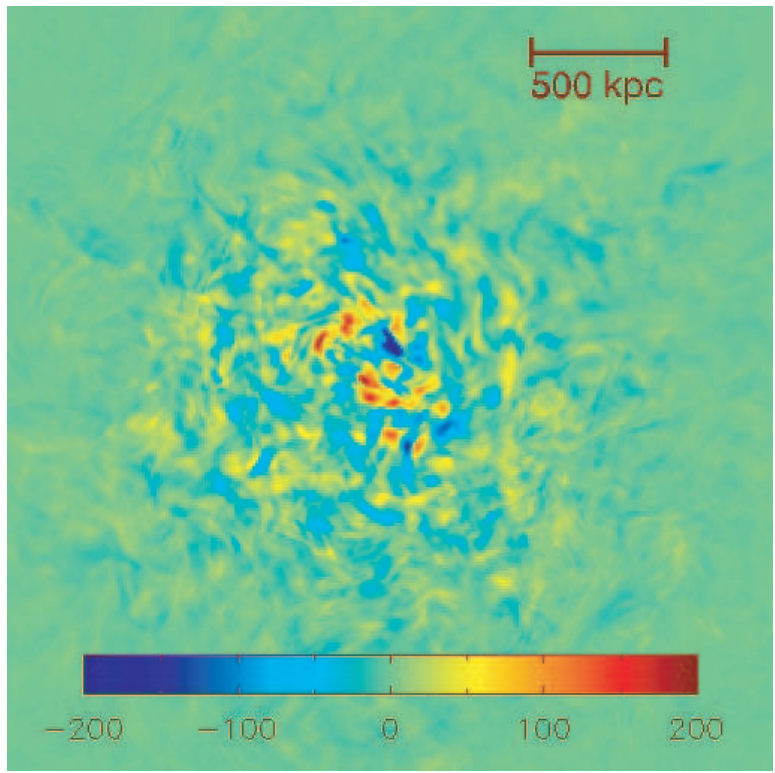,height=4in,width=4in} 
\end{center}
\caption{Faraday rotation measurement of the ICM by integrating
  to the mid-plane of the cluster. It covers a region of $2.86$ Mpc  
  $\times$ $2.86$
  Mpc (comoving) at z=0. The color range shown is from $-200$
  (blue) to $200$ (red) rad m$^{-2}$. The peak value of rotation
  measurement is about $\pm 400$ rad m$^{-2}$. Filamentary structures
  are quite common.
  \label{fig:rm}}
\end{figure}

We have also computed the synthetic FRM by integrating to the
mid-plane of the cluster. Fig. \ref{fig:rm} shows the spatial
distribution of FRM at $z=0$. The typical value of FRM is $\pm 200$
rad m$^{-2}$, with high values concentrated in the cluster core
region. Interestingly, the FRM map not only shows the small scale
variations reminiscent of the ICM MHD turbulence, but also displays
long, narrow filaments with dimensions of $300$ kpc $\times$ $90$ kpc.
The FRM magnitudes and spatial distributions from simulations are
quite consistent with observations of radio galaxies in clusters
\citep{Guidetti08,Eilek02,Taylor93}.

\section{Discussions}

The results presented here need to be taken as an initial step in
better understanding the evolution of AGN magnetic fields over the
lifetime of a cluster. The fact that the AGN magnetic fields are injected
very early in the history of the cluster formation (i.e., z=3), before the major
merger events, could be important. For AGN fields injected relatively late in the
cluster formation \citep[e.g.,][]{Xu08a} or injected into relatively quite
background ICM \citep[e.g.,][]{Liu08}, the magnetic fields may not
experience extensive
stretching and transport so that they could survive in a bubble
morphology without undergoing significant mixing with the background
ICM. 

Note that even though we have solved the ideal MHD equations, there is
clearly  numerical diffusion that has allowed the magnetic fields to
diffuse in the ICM. The rate of diffusion is often related to the
numerical Reynolds number and magnetic Reynolds number. We estimated
that these  numbers in our simulations are on the order of a few
hundred (this is  especially true in the cluster core region where
most of the  magnetic field energy resides). The small scale dynamo
theory and  simulations \citep{Boldyrev04} have shown that the dynamo
will grow  under such conditions, which is consistent with our
findings. The real  ICM could have Reynolds number as low as ten
\citep{Reynolds05},  but its magnetic Reynolds number is less well
determined,  especially in a magnetized turbulent medium. 

Since AGNs are commonly observed in galaxy clusters, one important
implication of our studies is that the magnetic fields from AGNs alone
are perhaps enough to ``seed'' the cluster and the ICM turbulence
generated and maintained by mergers will amply and spread the AGN
fields via a dynamo process. This is consistent with observations that
clusters with large scale radio emissions are often correlated with
cluster mergers \citep{Feretti05}.

\acknowledgments
  H. Li thanks F. Cattaneo and S. Colgate for discussions. 
  We thank the referee whose comments helped to improve the presentation
  and the quality of this Letter.
  This work was supported by the LDRD and IGPP programs at LANL. Computations
  were performed using the institutional computing resources at LANL.
  ENZO is developed at the Laboratory for Computational Astrophysics,
  UCSD with partial support from NSF grant ASST-0708960 to M.L.N.

\bibliographystyle{apj}

\begin{thebibliography}{31}
\expandafter\ifx\csname natexlab\endcsname\relax\def\natexlab#1{#1}\fi

\bibitem[{{Bernet} {et~al.}(2008){Bernet}, {Miniati}, {Lilly}, {Kronberg}, \&
  {Dessauges-Zavadsky}}]{Bernet08}
{Bernet}, M.~L., {Miniati}, F., {Lilly}, S.~J., {Kronberg}, P.~P., \&
  {Dessauges-Zavadsky}, M. 2008, \nat, 454, 302

\bibitem[{{Boldyrev} \& {Cattaneo}(2004)}]{Boldyrev04}
{Boldyrev}, S., \& {Cattaneo}, F. 2004, Phys. Rev. Lett., 92, 144501

\bibitem[{{Brandenburg} \& {Subramanian}(2005)}]{Brandenburg05}
{Brandenburg}, A., \& {Subramanian}, K. 2005, \physrep, 417, 1

\bibitem[{{Burbidge}(1959)}]{Burbidge59}
{Burbidge}, G.~R. 1959, \apj, 129, 849

\bibitem[{{Carilli} \& {Taylor}(2002)}]{Carilli02}
{Carilli}, C.~L., \& {Taylor}, G.~B. 2002, \araa, 40, 319

\bibitem[{{Colgate} \& {Li}(2000)}]{Colgate00}
{Colgate}, S.~A., \& {Li}, H. 2000, in IAU Symposium, Vol. 195, Highly
  Energetic Physical Processes and Mechanisms for Emission from Astrophysical
  Plasmas, ed. P.~C.~H. {Martens}, S.~{Tsuruta}, \& M.~A. {Weber} (Dordrecht: Kluwer), 255

\bibitem[{{Croston} {et~al.}(2005){Croston}, {Hardcastle}, {Harris}, {Belsole},
  {Birkinshaw}, \& {Worrall}}]{Croston05}
{Croston}, J.~H., {Hardcastle}, M.~J., {Harris}, D.~E., {Belsole}, E.,
  {Birkinshaw}, M., \& {Worrall}, D.~M. 2005, \apj, 626, 733

\bibitem[{{Dolag} {et~al.}(2002){Dolag}, {Bartelmann}, \& {Lesch}}]{Dolag02}
{Dolag}, K., {Bartelmann}, M., \& {Lesch}, H. 2002, \aap, 387, 383

\bibitem[{{Donnert} {et~al.}(2009){Donnert09}, {Dolag}, {Lesch}, \&
  {M{\"u}ller}}]{Donnert08}
{Donnert}, J., {Dolag}, K., {Lesch}, H., \& {M{\"u}ller}, E. 2009, \mnras, 392,
  1008

\bibitem[{{Dubois} \& {Teyssier}(2008)}]{Dubois08}
{Dubois}, Y., \& {Teyssier}, R. 2008, \aap, 482, L13

\bibitem[{{Eilek} \& {Owen}(2002)}]{Eilek02}
{Eilek}, J.~A., \& {Owen}, F.~N. 2002, \apj, 567, 202

\bibitem[{{Feretti}(1999)}]{Feretti99}
{Feretti}, L. 1999, in Diffuse Thermal and Relativistic Plasma in Galaxy
  Clusters, ed. H.~{Boehringer}, L.~{Feretti}, \& P.~{Schuecker} (Garching : MEP), 3

\bibitem[{{Feretti}(2005)}]{Feretti05}
{Feretti}, L. 2005, Adv. Space Res., 36, 729

\bibitem[{{Ferrari} {et~al.}(2008){Ferrari}, {Govoni}, {Schindler}, {Bykov}, \&
  {Rephaeli}}]{Ferrari08}
{Ferrari}, C., {Govoni}, F., {Schindler}, S., {Bykov}, A.~M., \& {Rephaeli}, Y.
  2008, Space Sci. Rev., 134, 93

\bibitem[{{Furlanetto} \& {Loeb}(2001)}]{Furlanetto01}
{Furlanetto}, S.~R., \& {Loeb}, A. 2001, \apj, 556, 619

\bibitem[{{Govoni} {et~al.}(2006){Govoni}, {Murgia}, {Feretti}, {Giovannini},
  {Dolag}, \& {Taylor}}]{Govoni06}
{Govoni}, F., {Murgia}, M., {Feretti}, L., {Giovannini}, G., {Dolag}, K., \&
  {Taylor}, G.~B. 2006, \aap, 460, 425

\bibitem[{{Guidetti} {et~al.}(2008){Guidetti}, {Murgia}, {Govoni}, {Parma},
  {Gregorini}, {de Ruiter}, {Cameron}, \& {Fanti}}]{Guidetti08}
{Guidetti}, D., {Murgia}, M., {Govoni}, F., {Parma}, P., {Gregorini}, L., {de
  Ruiter}, H.~R., {Cameron}, R.~A., \& {Fanti}, R. 2008, \aap, 483, 699

\bibitem[{{Kronberg} {et~al.}(2001){Kronberg}, {Dufton}, {Li}, \&
  {Colgate}}]{Kronberg01}
{Kronberg}, P.~P., {Dufton}, Q.~W., {Li}, H., \& {Colgate}, S.~A. 2001, \apj,
  560, 178

\bibitem[{{Kulsrud} {et~al.}(1997){Kulsrud}, {Cen}, {Ostriker}, \&
  {Ryu}}]{Kulsrud97}
{Kulsrud}, R.~M., {Cen}, R., {Ostriker}, J.~P., \& {Ryu}, D. 1997, \apj, 480,
  481

\bibitem[{{Li} {et~al.}(2006){Li}, {Lapenta}, {Finn}, {Li}, \&
  {Colgate}}]{Li06}
{Li}, H., {Lapenta}, G., {Finn}, J.~M., {Li}, S., \& {Colgate}, S.~A. 2006,
  \apj, 643, 92

\bibitem[{{Liu} {et~al.}(2008){Liu}, {Li}, {Li}, \& {Hsu}}]{Liu08}
{Liu}, W., {Li}, H., {Li}, S., \& {Hsu}, S.~C. 2008, \apjl, 684, L57

\bibitem[{{McNamara} \& {Nulsen}(2007)}]{McNamara07}
{McNamara}, B.~R., \& {Nulsen}, P.~E.~J. 2007, \araa, 45, 117

\bibitem[{{McNamara} {et~al.}(2005){McNamara}, {Nulsen}, {Wise}, {Rafferty},
  {Carilli}, {Sarazin}, \& {Blanton}}]{McNamara05}
{McNamara}, B.~R., {Nulsen}, P.~E.~J., {Wise}, M.~W., {Rafferty}, D.~A.,
  {Carilli}, C., {Sarazin}, C.~L., \& {Blanton}, E.~L. 2005, \nat, 433, 45

\bibitem[{{Nakamura} {et~al.}(2006){Nakamura}, {Li}, \& {Li}}]{Nakamura06}
{Nakamura}, M., {Li}, H., \& {Li}, S. 2006, \apj, 652, 1059

\bibitem[{{Reynolds} {et~al.}(2005){Reynolds}, {McKernan}, {Fabian}, {Stone},
  \& {Vernaleo}}]{Reynolds05}
{Reynolds}, C.~S., {McKernan}, B., {Fabian}, A.~C., {Stone}, J.~M., \&
  {Vernaleo}, J.~C. 2005, \mnras, 357, 242

\bibitem[{{Roettiger} {et~al.}(1999){Roettiger}, {Stone}, \&
  {Burns}}]{Roettiger99}
{Roettiger}, K., {Stone}, J.~M., \& {Burns}, J.~O. 1999, \apj, 518, 594

\bibitem[{{Ryu} {et~al.}(2008){Ryu}, {Kang}, {Cho}, \& {Das}}]{Ryu08}
{Ryu}, D., {Kang}, H., {Cho}, J., \& {Das}, S. 2008, Science, 320, 909

\bibitem[{{Taylor} \& {Perley}(1993)}]{Taylor93}
{Taylor}, G.~B., \& {Perley}, R.~A. 1993, \apj, 416, 554

\bibitem[{{Vogt} \& {En{\ss}lin}(2003)}]{Vogt03}
{Vogt}, C., \& {En{\ss}lin}, T.~A. 2003, \aap, 412, 373

\bibitem[{{Voit}(2005)}]{Voit05}
{Voit}, G.~M. 2005, Rev. Mod. Phys., 77, 207

\bibitem[{{Xu} {et~al.}(2008){Xu}, {Li}, {Collins}, {Li}, \& {Norman}}]{Xu08a}
{Xu}, H., {Li}, H., {Collins}, D., {Li}, S., \& {Norman}, M.~L. 2008, \apjl,
  681, L61

\end{thebibliography}

\clearpage

\end{document}